\documentclass[superscriptaddress,preprint,onecolumn,12pt]{revtex4}

\usepackage{amssymb}
\usepackage{amsmath}
\usepackage{graphicx}
\usepackage{color}
\usepackage{amsfonts}

\definecolor{red}{rgb}{1,0,0}

\begin{document}

\title{Grazing incidence fast atom diffraction for He atoms impinging on a
Ag(110) surface}
\author{C. A. R\'{\i}os Rubiano}
\affiliation{Instituto de Astronom\'{\i}a y F\'{\i}sica del Espacio (CONICET-UBA),
Casilla de correo 67, sucursal 28, 1428 Buenos Aires, Argentina.}
\author{G.A. Bocan}
\affiliation{Centro At\'{o}mico Bariloche, Comisi\'{o}n Nacional de Energ\'{\i}a At\'{o}%
mica, and Consejo Nacional de Investigaciones Cient\'{\i}ficas y T\'{e}%
cnicas, S.C. de Bariloche, R\'{\i}o Negro, Argentina.}
\author{M.S. Gravielle\thanks{%
Author to whom correspondence should be addressed.\newline
Electronic address: msilvia@iafe.uba.ar}}
\affiliation{Instituto de Astronom\'{\i}a y F\'{\i}sica del Espacio (CONICET-UBA),
Casilla de correo 67, sucursal 28, 1428 Buenos Aires, Argentina.}
\author{N. Bundaleski}
\affiliation{CeFITec, Department of Physics, Faculdade de Ci\^{e}ncias e Tecnologia,
Universidade Nova de Lisboa, P-2829-516, Caparica, Portugal.}
\author{H. Khemliche}
\affiliation{Institut des Sciences Mol\'{e}culaires d'Orsay, ISMO, Unit\'{e} mixte de
recherches CNRS-Universit\'{e} Paris-Sud, UMR 8214, B\^{a}timent 351,
Universit\'{e} Paris-Sud, 91405 Orsay Cedex, France.}
\author{P. Roncin}
\affiliation{Institut des Sciences Mol\'{e}culaires d'Orsay, ISMO, Unit\'{e} mixte de
recherches CNRS-Universit\'{e} Paris-Sud, UMR 8214, B\^{a}timent 351,
Universit\'{e} Paris-Sud, 91405 Orsay Cedex, France.}

\begin{abstract}
Experimental diffraction patterns produced by grazing scattering of fast
helium atoms from a Ag(110) surface are used as a sensitive tool to test
both the scattering and the potential models. To describe the elastic
collision process we employ the surface eikonal (SE) approximation, which \
is a semi-classical method that includes the quantum interference between
contributions coming from different projectile paths. The projectile-surface
potential is derived from an accurate density-functional theory (DFT)
calculation that takes into account the three degrees of freedom of the
incident projectile. A fairly good agreement between theoretical and
experimental momentum distributions is found for incidence along different
low-indexed crystallographic directions.
\end{abstract}

\maketitle
\date{\today }

\section{INTRODUCTION}

Since the unexpected observation of grazing incidence diffraction of fast
atoms (GIFAD) on crystal surfaces \cite{Schuller07,Rousseau07}, extensive
research, both experimental and theoretical, has been devoted to the subject
\cite%
{Schuller08,Manson08,Aigner08,Gravielle08,Bundaleski08,Schuller09PRB,Diaz09,Winter11}%
. The first experimental evidences of this phenomenon were reported at
insulator materials \cite{Schuller07,Rousseau07}, where the presence of a
wide band-gap helps to suppress inelastic processes, thus preventing quantum
decoherence \cite{Lienemann11}. Soon afterwards the effect was observed at
semi-conductor \cite{Khemliche09} and metallic surfaces \cite%
{Bundaleski08,Busch09} even though, in the case of metals, energy loss
values were found to be significant \cite%
{Bundaleski08,KhemlicheNIMB09,Bundaleski11}. In addition, GIFAD patterns
have displayed an exceptional sensitivity to the projectile-surface
interaction, making it possible to study very subtle contributions, like the
ones produced by surface rumpling \cite{Aigner08,Schuller10,Schuller12} or
by adsorbed structures \cite{Schuller09PRL,Seifert12}. Nowadays GIFAD is
becoming a promissory tool for examining the electronic and morphological
characteristics of solid-vacuum interfaces \cite%
{Khemliche09,Winter09,Momeni10,Seifert10}. \ \ \ \ \ \ \ \

The aim of this work is to investigate the diffraction patterns produced by
fast He atoms grazingly impinging on a Ag(110) surface. Since this collision
system corresponds to the first and simplest metallic case for which GIFAD
effects were experimentally observed \cite{Bundaleski08}, it provides a
useful prototype to test both the theoretical method and the surface
potential model. To describe the scattering process we employ a distorted
wave theory -- the surface eikonal (SE) approximation -- that makes use of
the eikonal wave function to represent the elastic collision with the
surface, while the projectile motion is classically described using
different initial conditions \cite{Gravielle08}. The SE approach has been
used to evaluate GIFAD\ distributions from insulator surfaces, providing
results in good agreement with the experimental data \cite%
{Gravielle09,SchullerPRA09,Gravielle11}. It has also been applied to the
elastic scattering of fast N atoms from a (111) silver surface, for which
asymmetries in the diffraction patterns might be originated by second atomic
layer effects in the surface potential \cite{Gravielle10}.

Due to the strong dependence of the interference patterns on the
atom-surface interaction, a crucial issue of the theoretical description is
the detailed representation of the projectile-surface potential. In Refs.
\cite{Bundaleski08,KhemlicheNIMB09} the He-Ag(110) potential was simulated
as a two-dimensional sinusoidal function, whose corrugation amplitude was
derived from experimental data by means of the Hard-Wall approximation. Here
we use a potential energy surface (PES) that was built from a large set \ of
\textit{ab initio} data obtained with the DFT-based \textquotedblleft
\textsc{Quantum Espresso}\textquotedblright\ code \cite{Giannozzi09},
combined with a sophisticated interpolation technique \cite{CRP}. From such
\textit{ab initio} values we derived a three-dimensional (3D) PES, taking
into account the projectile's three degrees of freedom. No average of the
surface potential along the incidence direction was considered in the
calculation.

In this article, eikonal projectile distributions derived by using the DFT
potential are compared with experimental data for three different incidence
directions:\ $[1\overline{1}0]$, $[001]$, and $[1\overline{1}2]$. The paper
is organized as follows. The theoretical method, including details of the
potential calculation, is summarized in Sec. II, results are presented and
discussed in Sec. III, and in Sec. IV we outline our conclusions. Atomic
units (a.u.) are used unless otherwise stated.\medskip

\section{THEORETICAL MODEL}

\subsection{The transition amplitude}

Within the SE approximation, the scattering state of the projectile is
represented with the eikonal wave function \cite{Gravielle08},
\begin{equation}
\Psi _{i}^{+}(\vec{R}_{P},t)=\phi _{i}(\vec{R}_{P})\exp (-i\eta (t)),
\label{fi-E}
\end{equation}%
where $\vec{R}_{P}$ is the position vector of the incident atom, $\phi _{i}(%
\vec{R}_{P})=(2\pi )^{-3/2}\exp (i\vec{K}_{i}\cdot \vec{R}_{P})$ is the
initial unperturbed wave function, with $\vec{K}_{i}$ the initial projectile
momentum, and the sign $+$ \ indicates the outgoing asymptotic conditions.
In Eq. (\ref{fi-E}) $\ $the function $\eta (t)$ denotes the eikonal-Maslov
phase, which depends on the classical position of the projectile at the time
$t$, $\vec{\mathcal{R}}_{P}(t)$, as \cite{SchullerPRA09}:
\begin{equation}
\eta (t)=\int_{-\infty }^{t}dt^{\prime }V_{SP}(\vec{\mathcal{R}}%
_{P}(t^{\prime }))+\phi _{M},  \label{Maslov}
\end{equation}%
where $V_{SP}$ is the projectile-surface interaction and $\phi _{M}=\nu \pi
/2$ \ is the Maslov correction that takes into account the phase change of
the scattering wave function as it passes through a focus, with the Maslov
index \ $\nu $ defined as in Ref. \cite{Avrin94}.

By introducing Eq. (\ref{fi-E}) in the usual definition of the T-matrix
element \cite{Joachain}, the SE transition matrix per unit area $\mathcal{A}$
reads \cite{SchullerPRA09}.
\begin{equation}
T_{if}^{(SE)}=\frac{1}{\mathcal{A}}\int\limits_{\mathcal{A}}d\vec{R}_{os}\
a_{if}(\vec{R}_{os}),  \label{T-eik}
\end{equation}%
where $\vec{R}_{os}$ is the initial position of the projectile on the
surface plane and

\begin{eqnarray}
a_{if}(\vec{R}_{os}) &=&\frac{1}{(2\pi )^{3}}\int\limits_{-\infty }^{+\infty
}dt\ |v_{z}(\vec{\mathcal{R}}_{P})|\ V_{SP}(\vec{\mathcal{R}}_{P})  \notag \\
&&\times \exp [-i\overrightarrow{Q}\cdot \vec{\mathcal{R}}_{P}-i\eta (t)]
\label{A-eik}
\end{eqnarray}%
is the transition amplitude associated with the classical path $\vec{%
\mathcal{R}}_{P}=\vec{\mathcal{R}}_{P}(\vec{R}_{os},t)$. The vector $%
\overrightarrow{Q}=\vec{K}_{f}-\vec{K}_{i}$ is the momentum transfer, with
the final projectile momentum $\vec{K}_{f}$ verifying energy conservation,
i.e. $\left\vert \vec{K}_{f}\right\vert =\left\vert \vec{K}_{i}\right\vert $%
. The function $v_{z}(\vec{\mathcal{R}}_{p})$ represents the component of
the projectile velocity that is perpendicular to the surface plane, with $%
\widehat{z}$ along the surface normal, aiming towards the vacuum region.

From Eq. (\ref{T-eik}), the differential probability, per unit of surface
area, for elastic scattering with final momentum $\vec{K}_{f}$ in the
direction of the solid angle $\Omega _{f}\equiv (\theta _{f},\varphi _{f})$
is $dP^{(SE)}/d\Omega _{f}=(2\pi )^{4}m_{P}^{2}\left\vert
T_{if}^{(SE)}\right\vert ^{2}$, where $m_{P}$ is the projectile mass, and $%
\theta _{f}$ and $\varphi _{f}$ are the final polar and azimuthal angles,
respectively, with $\varphi _{f}$ measured with respect to the incidence
direction in the surface plane. Details are given in Refs. \cite%
{Gravielle08,SchullerPRA09}.

\subsection{Projectile-surface interaction}

The interaction energy of the He atom with the Ag$(110)$ surface is
described with a full adiabatic 3D PES that depends on the atomic position $%
\vec{R}_{P}=(X,Y,Z)$. The PES is constructed from a grid of \textit{ab initio%
} energies for 42 $Z$ values and 6 ($XY$) sites, chosen as indicated in Fig.
1, over which an interpolation is performed \cite{CRP}.

All \textit{ab initio} data are obtained from the DFT-based
\textquotedblleft \textsc{Quantum Espresso}\textquotedblright\ code ~\cite%
{Giannozzi09}. The values of relevant input parameters are chosen so that
\textit{ab initio} energies are calculated to a prescribed accuracy
(differences $<5$ meV with respect to the converged result). The
exchange-correlation energy is calculated within the generalized gradient
approximation (GGA), using the Perdew-Burke-Ernzerhof energy functional
(PBE) ~\cite{PBE}. The electron-core interaction is described with
ultra-soft pseudopotentials ~\cite{US-pseudo}. The energy cutoff in the
plane-wave expansion is 35.0 Ryd for the wave functions and 245.0 Ryd for
the charge density and potential; the fractional occupancies are determined
through the broadening approach of Marzari-Vanderbilt \cite{Param1} with $%
\sigma =0.01$ Ryd.; and the Brillouin-zone integration is performed with a $%
10\times 7\times 1$ Monkhorst-Pack Grid of special $k$ points. The Ag
lattice constant is $a=7.865$ a.u.

The Ag$(110)$ surface is modeled by means of the supercell-slab scheme. A
four-layer slab is used with a $2\times 2$ cell in the plane parallel to the
surface and a supercell of length 25.03 a.u. along the normal to the surface
($\hat{z}$ axis). The energy for He-Ag(110) with the He atom midway between
slabs provides a reasonable representation of the asymptotic region and is
chosen as the energy reference. The surface interlayer distance is relaxed
from its bulk value $d_{0}=2.781$ a.u., keeping the two bottom layers fixed.
Geometry corrections due to relaxation amount to -9.14\% and +4.11\% for the
first and second interlayer distances respectively, in accord with
experimental results ~\cite{Relax}. Once relaxed, the slab is kept frozen
for the calculations that follow.

Given the closed-shell electronic structure of the He atom ($1s^{2}$), we
perform a non spin-polarized calculation of the ground state. A quality
check of the interpolation shows that the error introduced is $<1$ meV, well
below the prescribed accuracy for the \textit{ab initio} data. In Figs. \ref%
{fig:0} a) and b) we show the equipotential curves along the directions $%
[001]$ and $[1\bar{1}0]$, respectively, both starting from a site
corresponding to an atom of the first layer.

\section{RESULTS}

In this work we use the SE approximation to study momentum distributions of
\ $^{3}$He atoms elastically scattered from a Ag(110) surface under grazing
incidence conditions. Within the SE approach the perpendicular momentum
distribution is derived from the double differential probability $%
dP^{(SE)}/d\Omega _{f}$ as

\begin{equation}
\frac{dP^{(SE)}}{dQ_{\text{tr}}}=\int d\theta _{f}\ \frac{\cos \theta _{f}}{%
\sqrt{K_{fs}^{2}-Q_{\text{tr}}^{2}}}\frac{dP^{(SE)}}{d\Omega _{f}},
\label{dPdQperp}
\end{equation}%
where $K_{fs}=K_{f}\cos \theta _{f}$ is the final momentum parallel to the
surface and $Q_{\text{tr}}=K_{f}\cos \theta _{f}\sin \varphi _{f}$ is the\
component of the momentum transfer in the transverse direction, which is
perpendicular to the incidence direction on the surface plane.

\begin{figure}
\includegraphics[width=0.7\textwidth]{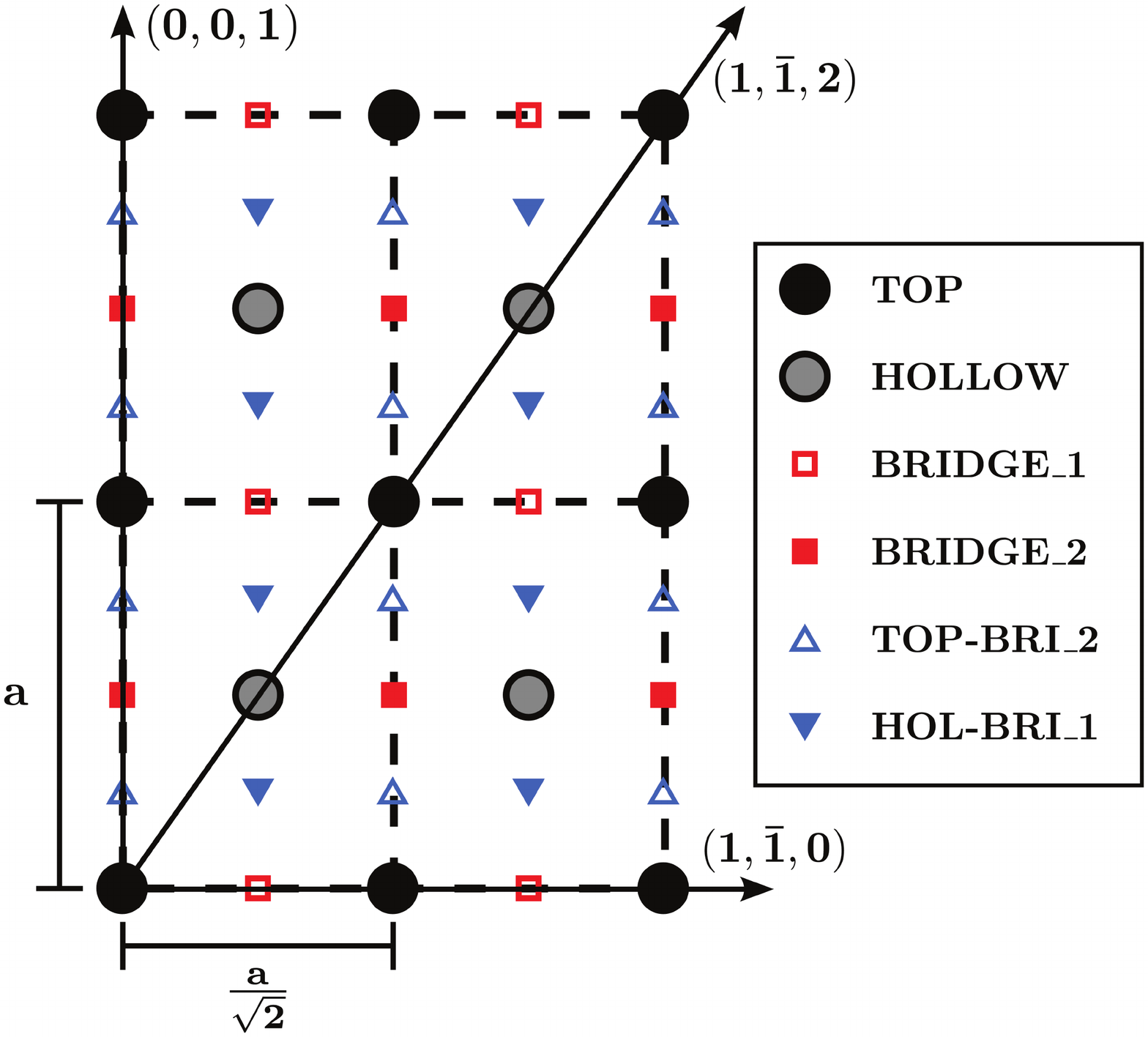}
\caption{(Color online) Geometry of the Ag(110) surface. $a$ is the lattice
constant. The six (XY) sites shown in the figure correspond to the ones used
for the PES calculation. Circles, TOP and HOLLOW sites standing for atoms of
the first and second layers, respectively; squares, BRIDGE\_1 and \
BRIDGE\_2 sites corresponding to the middle points between first and second
TOP atomic neighbors, respectively; triangles, middle points between the
mentioned sites. The incidence directions of He atoms are also indicated.}
\label{fig:1}
\end{figure}

Like in Refs. \cite{Bundaleski08,KhemlicheNIMB09}, projectile impact along
three different channels of the silver surface - $[1\bar{1}0]$, $[001]$, and
$[1\bar{1}2]$ \ - is considered (see Fig. \ref{fig:1}). Note that in
contrast with the notation employed in Refs. \cite%
{Bundaleski08,KhemlicheNIMB09}, here the figures will be labeled with the
direction of the incident beam, rather than with the probed direction, which
is perpendicular to $\vec{K}_{i}$. For each of these directions, we
evaluated $T_{if}^{(SE)}$ from Eq. (\ref{T-eik}) using $4\times 10^{5}$
classical trajectories with random initial positions $\vec{R}_{os}$ that
vary within an area $\mathcal{A}$.  In this work, in order to resemble the
incident wave packet we use Gaussian distributions along the incidence and
transverse directions, with full widths at half maximum determined by the
size of $\mathcal{A}$, to evaluate the random initial positions of
projectile trajectories \cite{Joachain,random}. For an almost perfect
plane-wave beam, the area $\mathcal{A}$ is large and includes several
reduced unit cells.

In Fig. \ref{fig:2} we compare theoretical and experimental perpendicular
momentum spectra for He atoms impinging along the $[1\bar{1}0]$ direction.
The incidence energy is $E_{i}=K_{i}^{2}/(2m_{P})=500\ $eV and the polar
incidence angle, measured with respect to the surface plane, $\ $is $\theta
_{i}=0.75^{%
{{}^o}%
}$. It corresponds to a perpendicular energy, associated with the movement
normal to the surface, $E_{i\perp }=E_{i}\sin ^{2}\theta _{i}=86\ $meV.
Under this incidence condition, the experimental momentum distribution
presents a rich diffraction pattern, with maxima and minima almost
symmetrically placed with respect to the incidence direction, for which $Q_{%
\text{tr}}=0$. For a detailed analysis of this spectrum, SE differential
momentum probabilities, obtained by considering in Eq. (\ref{T-eik}) an
integration region $\mathcal{A}$ equal to $8\times 8$ reduced unit cells,
are displayed in the figure with a blue dashed line. These results present
narrow Bragg maxima placed at $Q_{\text{tr}}=m2\pi /d$, with $m$ an integer
number and $d$ the distance between parallel atomic rows along the incidence
channel, whose positions coincide with those of the experimental maxima.
Within the SE approach, the width of these Bragg peaks is governed by the
number of reduced unit cells along the perpendicular direction, $n_{\text{tr}%
}$ (in our case, $n_{\text{tr}}=8$\ ), decreasing as $n_{\text{tr}}$
increases, while their intensities are determined by the supernumerary
rainbow distribution, as explained below.

The two different mechanisms - Bragg diffraction and supernumerary rainbows
- that are present in GIFAD patterns \cite{Schuller08} can be analyzed
separately with the SE model. The SE transition matrix can be factorized as
\cite{Gravielle11}
\begin{equation}
T_{if}^{(SE)}=\widetilde{T}_{1}^{(SE)}S_{n_{\text{tr}}}(Q_{\text{tr}}),
\label{Tbragg}
\end{equation}%
where $\widetilde{T}_{1}^{(SE)}$ is derived from Eq. (\ref{T-eik}) by
evaluating the $\vec{R}_{os}$-integral over \textit{one} reduced unit cell,
and $S_{n_{\text{tr}}}(Q_{\text{tr}})=\sin (n_{\text{tr}}\ \beta )/\left( n_{%
\text{tr}}\sin \beta \right) $, with $\beta =Q_{\text{tr}}d/2$, for
incidence along the $[1\bar{1}0]$ and $[001]$ directions. \ The factor $%
\widetilde{T}_{1}^{(SE)}$, associated with supernumerary rainbows, is
produced by interference between projectile trajectories whose initial
positions are separated by a distance smaller than $d$. In turn, $S_{n_{%
\text{tr}}}(Q_{\text{tr}})$, which gives rise to the Bragg peaks, is due to
interference between projectile trajectories whose initial positions are
separated by a distance just equal to a multiple of $d$. The positions of
Bragg peaks, indicated with vertical dashed lines in Figs. 3 and 4, provide
information about the crystallographic structure only, \ but their
intensities, which are modulated by the supernumerary rainbow factor $%
\widetilde{T}_{1}^{(SE)}$, depend strongly on the shape of the PES across
the incidence channel. In fact, the factor $\widetilde{T}_{1}^{(SE)}$
completely determines the number and the intensity of observed Bragg maxima,
even suppressing them, as approximately happens for the Bragg peaks of order
$m=\pm 1$ in Fig. \ref{fig:2}.

In order to compare with the experimental data, in Fig. \ref{fig:2} we also
plot SE differential probabilities convoluted with a Lorentz function (red
solid line) to simulate the experimental conditions. The parameters of the
line broadening are taken from the observed linewidths, as stated in Ref.
\cite{Bundaleski08}. Such a convolution takes into account not only the
experimental divergence of the incident beam but also the broadening
introduced by both thermal vibrations of lattice atoms and inelastic
processes, which contribute to spoil the coherence \cite{background}. We
found a good agreement between the convoluted SE results and the experiment
in the whole range of perpendicular momenta, with the exception of the
maxima of order $m=\pm 3$ whose intensity is overestimated by the SE curve.
This fact is related to the sharpness of the rainbow peaks of the envelope
function $\widetilde{T}_{1}^{(SE)}$, which is originated by the classical
description of the projectile motion that does not include the finite
intensity on the dark side of the classical rainbow \cite{Berry72}. Such
classical rainbow maxima affect the intensity of the outermost Bragg peaks
of the GIFAD pattern only when they are close to each other, as observed in
the figure. However, this deficiency does not influence the SE distribution
at smaller transverse momenta.

\begin{figure}
\includegraphics[width=0.7\textwidth]{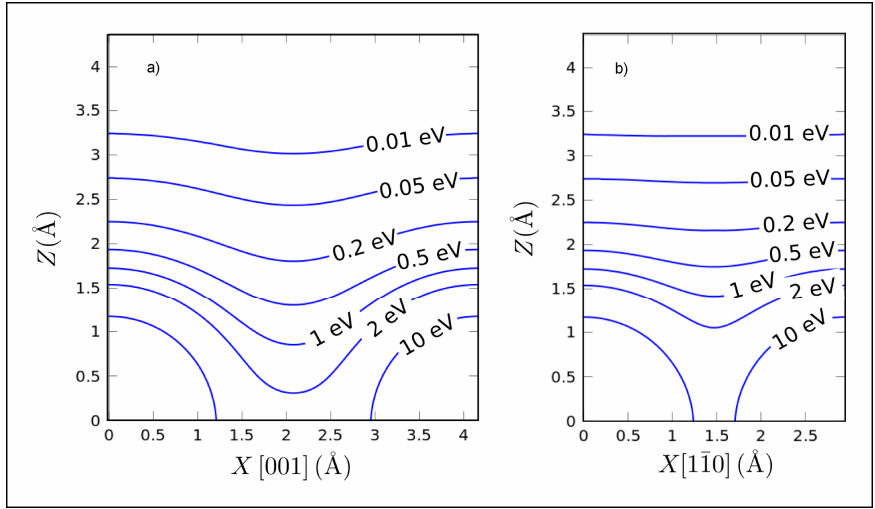}
\caption{(Color online) Distances Z to the surface plane (topmost atomic
layer) of equipotential curves for the interaction between the He atom and
the Ag(110) surface. The value X=0 corresponds to a TOP site, as indicated
in Fig. \protect\ref{fig:1}. \ a) Equipotential curves as a function of the
coordinate along the $[001]$ direction, $X[001]$, for $X[1\bar{1}0]=0$; b)
similar to a) as a function of the coordinate along the $[1\bar{1}0]$
direction, $X[1\bar{1}0]$, for $X[001]=0$.}
\label{fig:0}
\end{figure}

\begin{figure}
\includegraphics[width=0.7\textwidth]{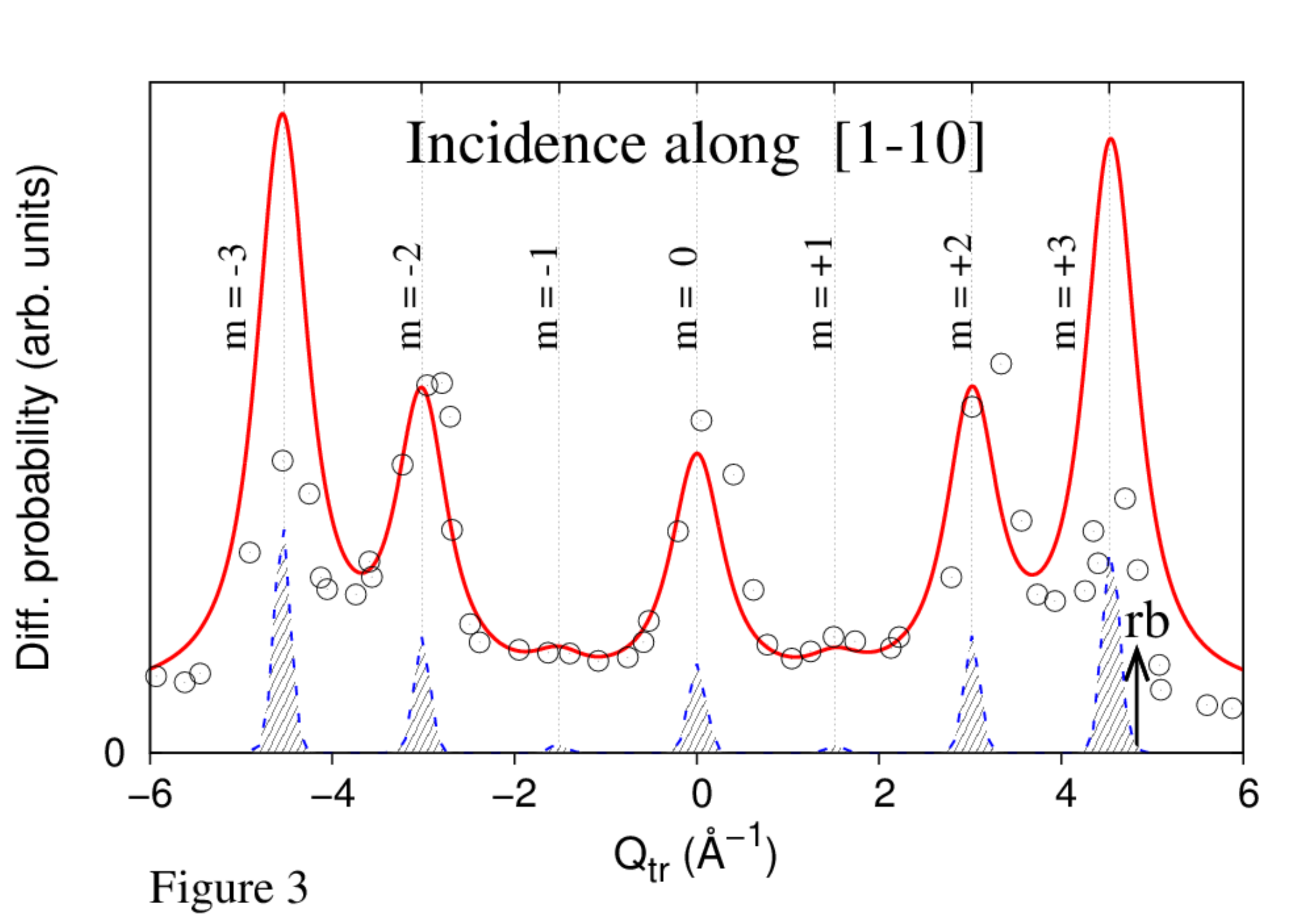}
\caption{(Color online) Momentum distribution, as a function of the
perpendicular momentum transfer $Q_{\text{tr}}$, for $^{3}$He atoms
impinging on Ag(110) surface along the $[1\bar{1}0]$ direction. The
incidence energy and angle are $E_{i}=500\ $eV and $\protect\theta %
_{i}=0.75^{{{}^{o}}}$, respectively. Solid red line, SE differential
momentum probability convoluted to include uncertainties as explained in the
text \protect\cite{background}; dashed blue line, SE differential momentum
probability for $n_{\text{tr}}=8$\ (without convolution); empty circles,
experimental data from Ref. \protect\cite{Bundaleski08}. The vertical dashed
lines show Bragg peak positions and the arrow indicates the position of the
classical rainbow maximum.}
\label{fig:2}
\end{figure}

\begin{figure}
\includegraphics[width=0.7\textwidth]{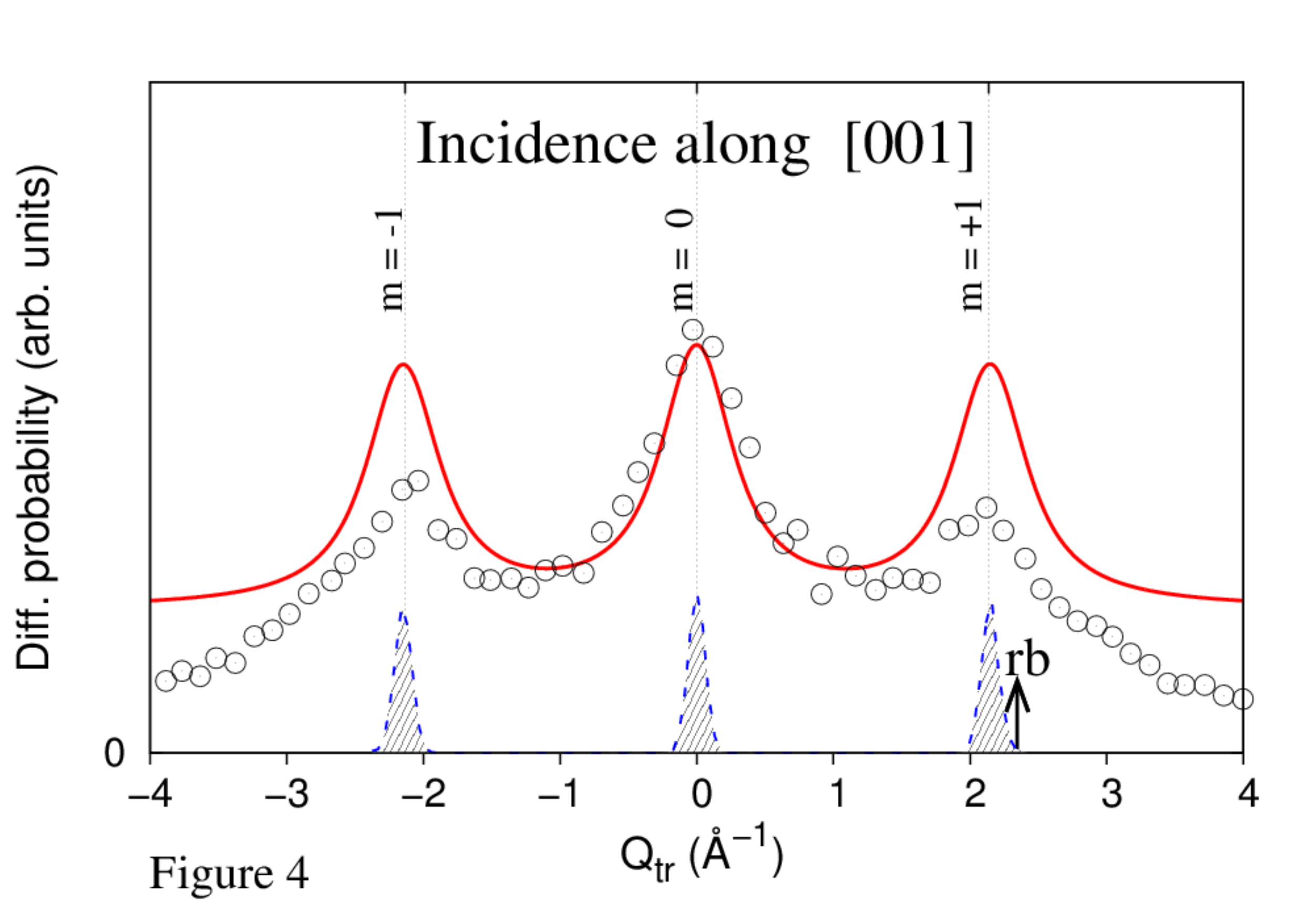}
\caption{ Similar to Fig. \protect\ref{fig:2} for $^{3}$He atoms impinging
along the [001] direction with $E_{i}=500\ $eV and $\protect\theta _{i}=1.0^{%
{{}^{o}}}$. Empty circles, experimental data extracted from Ref.
\protect\cite{Bundaleski11}. }
\label{fig:3}
\end{figure}

The final perpendicular momentum distribution of a He beam impinging along a
less corrugated crystallographic channel - the $[001]$ channel - is shown in
Fig. \ref{fig:3} for the incidence parameters $E_{i}=500\ $eV and $\theta
_{i}=1.0^{%
{{}^o}%
}$ (i.e., a perpendicular energy $E_{i\perp }=180\ $meV). Here, SE results
obtained by considering initial positions inside $n_{\text{tr}}=16$\ unit
cells are plotted in the figure together with\ the convoluted values. The SE
momentum spectrum, including experimental and inherent uncertainties through
convolution, is in fairly good accord with the experimental data. Notice
that the $Q_{\text{tr}}$ position of the classical rainbow peak depends on $%
E_{i\perp }$, and for this incidence condition it is again close to the
outermost Bragg maximum. On the other hand, the number of observed Bragg
maxima is determined by $\widetilde{T}_{1}^{(SE)}$, being sensitive to the
potential contour across the incidence direction. Then, as a consequence of
the much lower corrugation of the channel, observed in Fig. \ref{fig:0}, the
projectile momentum spectrum is narrower than the one of Fig. \ref{fig:2}.

\begin{figure}
\includegraphics[width=0.7\textwidth]{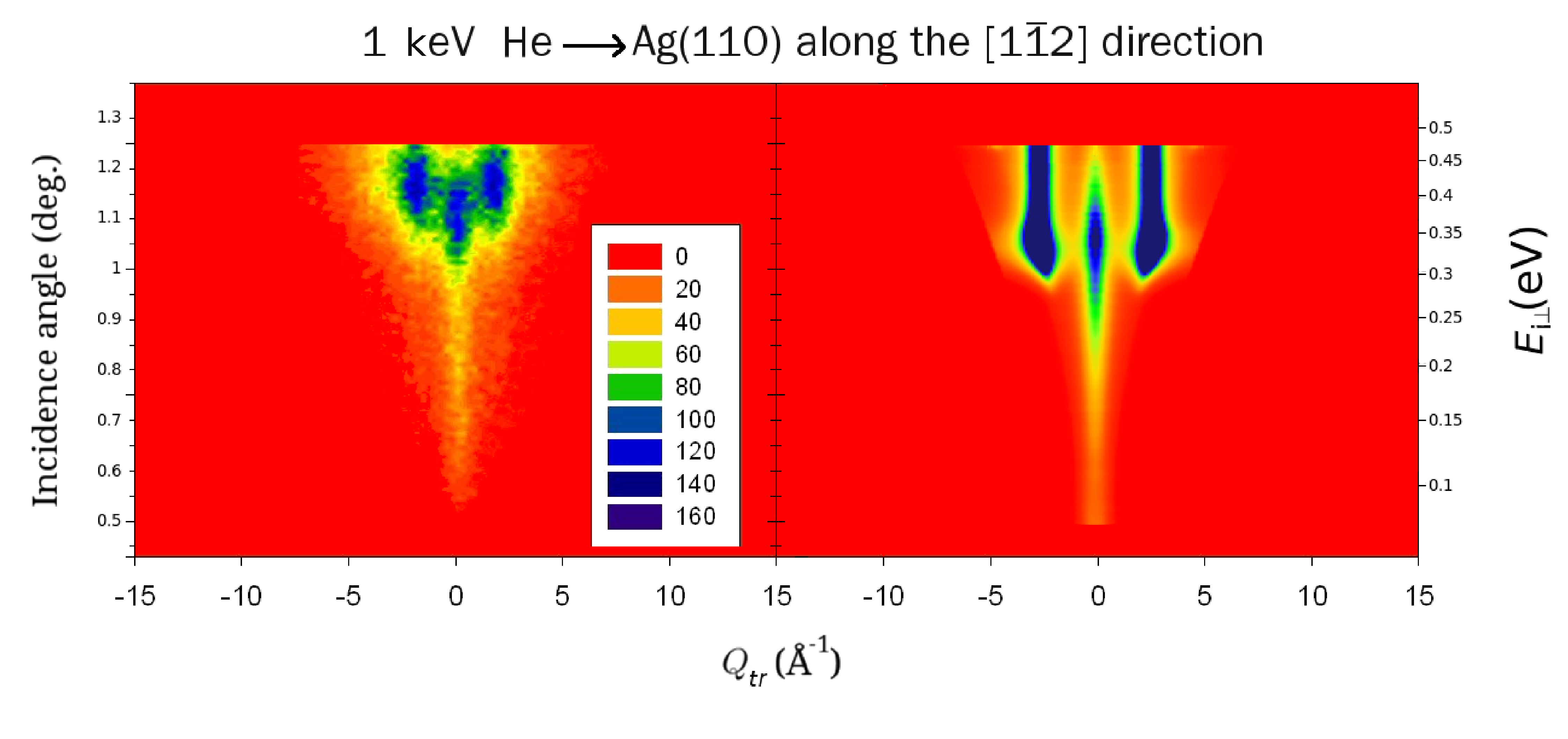}
\caption{ (Color online) Two-dimensional diffraction chart displaying (a)
experimental intensities and (b) SE differential momentum probabilities, as
a function of the transverse momentum transfer $Q_{\text{tr}}$ and the
incidence angle $\protect\theta _{i}$ (or the normal energy $E_{i\perp }$),
for $1$ keV $^{3}$He atoms impinging along the $[1\overline{1}2]$ direction.}
\label{fig:4}
\end{figure}

Finally, in Fig. \ref{fig:4} we compare experimental and theoretical
diffraction charts for $1\ $keV $^{3}$He atoms impinging on the silver
surface along the $[1\bar{1}2]$ channel. These diffraction charts display
the intensity of the projectile distribution as a function of the transverse
transferred momentum and the incidence angle $\theta _{i}$ ( or the normal
impact energy), providing an overall scenery of the GIFAD patterns for this
low-indexed crystallographic direction. The experimental diffraction chart
of Fig. \ref{fig:4} (a) was obtained by considering intensity distributions
corresponding to 38 different incidence angles. Like in Fig. \ref{fig:3}, as
a consequence of the low corrugation of the channel, only three diffraction
orders - $m=0$ and $m=\pm 1$ - are visible in the figure. However, since the
number of observed Bragg maxima depends strongly on the shape of the
equipotential curves across the incidence channel \cite{Aigner08}, and
different values of $E_{i\perp }$ allow one to probe such potential contours
for different distances to the surface, the general agreement between the
experimental and simulated two-dimensional diffraction charts of Fig. \ref%
{fig:4} is a signature of the quality of the present DFT potential in the $%
0.10$ - $0.45$ eV range of perpendicular energies.

\section{CONCLUSIONS}

GIFAD patterns for helium atoms colliding with a silver surface are used to
test both the SE approach and the \textit{ab initio }DFT surface potential,
for the case of a metal target. \ The SE approximation takes into account
the quantum interference produced by the coherent superposition of
transition amplitudes for different projectile paths that end with the same
final momentum, while the \textit{ab initio} potential was obtained from DFT
by making use of the \textquotedblleft \textsc{Quantum Espresso}%
\textquotedblright\ code.

For the He-Ag(110) system, experimental momentum spectra along different
crystallographic directions display defined interference structures that are
fairly well reproduced by the theoretical results. Taking into account the
extreme sensitivity of fast atom diffraction patterns, this agreement
between SE differential momentum probabilities and experimental data is
indicative of the proper description of the atom-surface interaction given
by the DFT potential.  Further experimental and theoretical research has
been recently carried out in order to make a deeper examination of the
potential model, considering a wider normal energy range \cite{Rios13}.

\begin{acknowledgments}
C.R.R and M.S.G. acknowledge financial support from CONICET, UBA, and ANPCyT
of Argentina. One of us (M.S.G) would like to express gratitude to the
Universit\'{e} Paris-Sud for support during a portion of this work. G.A.B.
acknowledges financial support by ANPCyT and is also thankful to Dr. H.F.
Busnengo, Dr. J.D. F\"{u}hr and Dr. M.L. Martiarena regarding the PES
calculation. The experimental work was supported by the ANR under contract
number ANR-07-BLAN-0160-01.
\end{acknowledgments}

\begin{thebibliography}{99}
\bibitem{random} When few reduced unit cells are considered in the
transverse direction, such a soft random function avoids spurious
interferences originated by the uniform random distribution, which are
associated with the one-slit problem.
\bibitem{background} In order to account for the incoherent background
reported in Ref. \cite{Bundaleski08}, a constant term was also added to the
convoluted SE probabilities.
\end{thebibliography}
\end{document}